\documentstyle[preprint,aps]{revtex}

\begin{document}

\draft
\title
{The spin-statistics connection, the Gauss-Bonnet theorem and 
the Hausdorff dimension of the quantum paths}

\author
{Wellington da Cruz\footnote{E-mail: wdacruz@exatas.uel.br}}

\address
{Departamento de F\'{\i}sica,\\
 Universidade Estadual de Londrina, Caixa Postal 6001,\\
Cep 86051-970 Londrina, PR, Brazil\\}
 
\date{\today}

\maketitle

\begin{abstract}

We obtain an explicit expression relating the writhing number, $W[C]$, of the quantum 
path, $C$, with any value of spin, $s$, of the particle which 
sweeps out that closed curve. We consider a fractal approach 
to the fractional spin particles and , in this way, we make clear a deeper 
connection between the 
Gauss-Bonnet theorem with the spin-statistics relation via 
the concept of Hausdorff dimension, $h$, associated to 
the fractal quantum curves of the particles:

\[
\frac{h}{2+2s}=W[C]=\frac{1}{4\pi}\oint_{C}d\;x_{\alpha}\oint_{C}d\;x_{\beta}\;
\varepsilon^{\alpha\beta\gamma}\;\frac{(x-y)_{\gamma}}{|x-y|^3}.
\]  
 
\end{abstract}


We have considered the fractal or Hausdorff 
dimension\footnote{The fractal dimension $h$ can be defined by

\[
h-1=\lim_{R\rightarrow 0}\frac{\ln\;(L/l)}{\ln\;(R)},
\]

 where $L$ is the perimeter of a closed curve, $C$, and
  $l$ is the usual length for the resolution $R$. The curve is 
 covering with $\frac{l}{R}$ spheres of diameter $R$ and so 
 a fractal curve is scale invariant, 
 self-similar and has a non-integer dimension. A fractal curve is 
continuous but not differentiable\cite{R2}.}, $h$, 
associated to the quantum paths of fractional spin particles\cite{R1} . These 
objects are classified in universal classes of fractons understood as charge-flux systems 
living in two-dimensional multiply connected space. The spin, $s$, of 
the particles are related with the Hausdorff dimension by $h=2-2s$, 
with $0 < s < \frac{1}{2}$. This expression is analogous to the fractal 
dimension of the graph of the functions in the context of the fractal 
geometry and given by $\Delta[C]=2-H$, 
with $ 0 < H <1$, where $H$ is the H\"older exponent\cite{R2} which appears 
in different scenarios of physical theories\cite{R3}. We have also 
established a fractal-deformed Heisenberg algebra for fractons 
which generalizes the fermionic and bosonic ones\cite{R4}. 
The fractal-deformed Heisenberg 
algebra is obtained of the relation

\begin{eqnarray}
{\bf a}(x){\bf a}^{\dagger}(y)-f[\pm h]{\bf a}^{\dagger}(y)
{\bf a}(x)=\delta (x-y),
\end{eqnarray}

\noindent between creation and annihilation operators. The factor of 
deformation\footnote{The plus sign of $h$ stands for anticlockwise 
exchange and the minus sign for clockwise exchange.} is defined as 

\begin{eqnarray}
f[\pm h]=exp\left(\pm\imath h\pi\right), 
\end{eqnarray}

\noindent such that for $h=1$ and $x=y$, we reobtain the fermionic 
anticommutation relations $\left\{{\bf a}(x),
{\bf a}^{\dagger}(x)\right\}=1$, and for  $h=2$ and $x=y$, we 
reobtain the bosonic 
commutation relations $\left[{\bf a}(x),
{\bf a}^{\dagger}(x)\right]=1$. If $x\neq y$ and 
$1$$\;$$ < $$\;$$h$$\;$$ <$$\;$$ 2$, we have nonlocal operators 
for fractons

\begin{eqnarray}
{\bf a}(x){\bf a}^{\dagger}(y)=f[\pm h]{\bf a}^{\dagger}(y)
{\bf a}(x).
\end{eqnarray}

\noindent The phase factor appears when we interchange two identical fractons and this 
phase is connected with the writhing number of the quantum path. On the other hand, 
the Gauss-Bonnet theorem\footnote{This theorem establishes a 
connection between a topological object, as the genus of the surface, and metrical 
entities, as distances and angles\cite{R5}.} which relates the Euler characteristics, $\chi$, 
of a closed surface, $S$, with the total Gaussian curvature, $K$, of that surface and given by

\begin{equation}
\label{e.5}
2-2g=\chi=\frac{1}{2\pi}\int_{S}K\;dA,
\end{equation}

\noindent where $g$ is the genus of the surface, has also a deeper relation 
with the spin-statistics connection . There exists a version 
of the Gauss-Bonnet formula which involves a topological invariant, 
the self-linking number of the curve\footnote{The objects $T[C]$ and $W[C]$ 
are metrical properties of the path and can assume any value, while $SL[C]$ 
must be an integer\cite{R5}.}, $SL[C]=T[C]+W[C]$, 
where $T[C]$ is the torsion and $W[C]$ is the writhing number of the closed curve $C$\cite{R5}. 
For $SL[C]=0$, $T[C]=-W[C]$ and this equivalence\footnote{The exponentiation of $SL[C]$, 
for a closed curve, confirmes this result given that topological invariant is an integer.} 
between the  torsion and the writhing number implies that the torsion of the quantum path 
of the particle is related to its spin\cite{R6,R5}.

The expectation value of one loop\cite{R6,R5} is given by

\begin{equation}
\langle\; exp\left(i\oint_{C}dx_{\mu}\;A^{\mu}(x)\right)\;\rangle=exp
\left(i\;\theta\;W[C]\right),
\end{equation}

\noindent where $\theta=2\pi s$ is the statistical parameter and the writhing number appears as

\begin{equation}
W[C]=\frac{1}{4\pi}\oint_{C}dx_{\alpha}\;\oint_{C}dx_{\beta}\;\varepsilon^{\alpha\beta\gamma}
\;\frac{(x-y)_{\gamma}}{|x-y|^3}.
\end{equation}

\noindent This way, the spin phase factor $exp\left(i\; 2\pi s\;W[C]\right)$ 
can be related with 
the factor of deformation, $exp(i\; h\pi)$, of the fractal-deformed Heisenberg algebra, 
taking into account the spin-statistics connection\footnote{Fractons 
satisfy this relation\cite{R1}.}, $\nu=2s$, and its 
periodicity $\nu\rightarrow \nu+2$, we obtain

\begin{equation}
\label{e.8}
\frac{h}{2+2s}=W[C]=\frac{1}{4\pi}\oint_{C}d\;x_{\alpha}\oint_{C}d\;x_{\beta}\;
\varepsilon^{\alpha\beta\gamma}\;\frac{(x-y)_{\gamma}}{|x-y|^3}.
\end{equation}  
 
\noindent Now, the geometrical parameter\footnote{This fractal parameter 
can be extracted from the propagators of the particles in the momentum space\cite{R7,R1}. 
We have also observed that the topological meaning of $\nu=
\theta/\pi=2s$ comes from $h$\cite{R1}.} 
$h$ encodes the fractal properties\footnote{The fractal character of the quantum 
path was noted by Feynman\cite{R8}, and this property 
reflects the Heisenberg uncertainty principle.} 
of the quantum path associated to the particle and 
in terms of its spin can be written as\cite{R1} 

\begin{eqnarray}
&&h=2-2s,\;\;\;\; 0 < s < \frac{1}{2};\;\;\;\;\;\;\;\;
 h=2s,\;
\;\;\;\;\;\;\;\;\;\;\;\; \frac{1}{2} < s < 1;\\
&&h=4-2s,\;\;\;\; 1 < s < \frac{3}{2};\;\;\;\;\;\;\;\;
 h=2s-2,\;
\;\;\;\;\;\; \frac{3}{2} < s < 2;\\ 
&&etc.\nonumber
\end{eqnarray}

\noindent Thus the writhing number associated to the path 
is written in terms of the spin\footnote{The spin can take any 
value as the writhing number.} of the particles

\begin{eqnarray}
&&W[C]=\frac{2-2s}{2+2s},\;\;\;\;\;\; 0 < s < \frac{1}{2};\;\;\;\;\;\;\;\;
 W[C]=\frac{2s}{2+2s},\;
\;\;\;\;\;\; \frac{1}{2} < s < 1;\\
&&W[C]=\frac{4-2s}{2+2s},\;\;\;\;\;\; 1 < s < \frac{3}{2};\;\;\;\;\;\;\;\;
 W[C]=\frac{2s-2}{2+2s},\;
\;\;\;\;\;\; \frac{3}{2} < s < 2;\\ 
&&etc.\nonumber
\end{eqnarray}

Here, we observe that our fractal approach to the fractional spin 
particles gives us a new perspective for such charge-flux systems, 
because we define universal classes of fractons\footnote{A symmetry of duality 
between these classes is also defined as $\tilde{h}=3-h$ and this implies 
another one, i. e., a fractal supersymmetry for pairs of particles 
$\left(s, s+\frac{1}{2}\right)$\cite{R1}.} labelled just by the 
Hausdorff dimension. This notion of set of particles with distinct values of spin 
is defined in the same way 
that fermions (bosons) constitute a universal class of particles 
with half-integer (integer) values of spin 
satisfying the Fermi-Dirac (Bose-Einstein) distribution function. 
Fractons as charge-flux 
systems satisfy a specific fractal distribution function\footnote
{We understanding this formula as a quantum-geometrical description 
of the statistical laws of nature.}\cite{R1}

\begin{eqnarray}
\label{e.h} 
n[h]=\frac{1}{{\cal{Y}}[\xi]-h},
\end{eqnarray}

\noindent which appears 
as a simple and elegant generalization of the fermionic, $h=1$,  
and bosonic, $h=2$, distributions for particles with 
braiding properties. The function 
${\cal{Y}}[\xi]$ satisfies the equation 

\begin{eqnarray}
\xi=\biggl\{{\cal{Y}}[\xi]-1\biggr\}^{h-1}
\biggl\{{\cal{Y}}[\xi]-2\biggr\}^{2-h},
\end{eqnarray}

\noindent with $\xi=\exp\left\{(\epsilon-\mu)/KT\right\}$.

The fractal von Neumann entropy per state in terms 
of the average occupation number is given as\cite{R1,R4}

\begin{eqnarray}
\label{e5}
{\cal{S}}[h,n]&=& K\left[\left[1+(h-1)n\right]\ln\left\{\frac{1+(h-1)n}{n}\right\}
-\left[1+(h-2)n\right]\ln\left\{\frac{1+(h-2)n}{n}\right\}\right],
\end{eqnarray}

\noindent and it is associated to the fractal distribution function Eq.(\ref{e.h}).

Finally, the formal aspect of the Eqs.(\ref{e.5},\ref{e.8}) 
is enough suggestive for other ideas.


\begin{thebibliography}{99}


\bibitem{R1} W. da Cruz, Int. J. Mod. Phys. {\bf A 15} (2000) 3805.
\bibitem{R2} C. Tricot, {\it Curves and Fractal Dimension} 
(Springer-Verlag, New York, 1995).
\bibitem{R3} K. M. Kolwankar and A. D. Gangal, Pramana J. Phys. {\bf 48} (1997) 49 ;\\ 
Chaos {\bf 6} (1996) 505 ;\\
P. Meakin, {\it Fractals, Scaling and Growth far from Equilibrium}\\
(Cambridge University Press, Cambridge, 1998).
\bibitem{R4} W. da Cruz, Physica {\bf A 313} (2002) 446 .
\bibitem{R5} C-H. Tze and S. Nam, Ann. Phys. {\bf 193} (1989) 419;\\
C-H. Tze, Int. J. Mod. Phys. {\bf A 3} (1988) 1959;\\
S. Forte, Rev. Mod. Phys. {\bf 64} (1992) 193;\\
M. D. Frank-Kamenetskij and A. V. Vologodskij, Sov. Phys. Usp. {\bf 24} (1981) 679;\\
and references therein.
\bibitem{R6} A. M. Polyakov, Mod. Phys. Lett. {\bf A 3} (1988) 325;\\
J. Grundberg {\it et al.}, Phys. Lett. {\bf B 218} (1989) 321.
\bibitem{R7} A. M. Polyakov, in {\it Proc. Les
 Houches Summer School
 {\bf vol. IL}},\\
  ed. E. Br\'ezin and J. Zinn-Justin
  (North Holland, 1990), p.305.
\bibitem{R8}  R. P. Feynman and A. R. Hibbs, {\it Quantum Mechanics and Path Integrals}\\ 
( MacGraw-Hill, New York, 1965 ), pp. 176-177.
\end{thebibliography}
\end{document}